\documentclass[sigconf,natbib=true,nonacm]{acmart}

\usepackage{enumitem}

\usepackage{tcolorbox}
\usepackage{hyperref}

\usepackage{graphicx}
\usepackage{float}
\usepackage{amsmath}
\usepackage{array}
\usepackage{graphicx}
\usepackage{subcaption}
\usepackage{overpic}

\usepackage{ifthen}
\usepackage{xcolor}

\newboolean{showcolorblock}
\setboolean{showcolorblock}{true}  

\newcommand{\anoni}[2]{#2}

\newcommand{\hidecontentstart}{\iffalse}%
\ifthenelse{\boolean{showcolorblock}}{%
    
}{%
    \excludecomment{colorblock}
}

\title{LLM-based Relevance Assessment Still Can't Replace
Human Relevance Assessment}

\author{Charles L. A. Clarke}
\affiliation{%
   \institution{University of Waterloo}
   \country{Canada}
}
\author{Laura Dietz}
\affiliation{%
   \institution{University of New Hampshire}
   \country{USA}
}

\begin{document}

\begin{abstract}

The use of large language models (LLMs) for relevance assessment in information retrieval has gained significant attention, with recent studies suggesting that LLM-based judgments provide comparable evaluations to human judgments. Notably, based on TREC 2024 data, \citet{upadhyay2024largescalestudyrelevanceassessments} make a bold claim that LLM-based relevance assessments, such as those generated by the {\sc Umbrela} system, can fully replace traditional human relevance assessments in TREC-style evaluations. This paper critically examines this claim, highlighting practical and theoretical limitations that undermine the validity of this conclusion. 

First, we question whether the evidence provided by \citet{upadhyay2024largescalestudyrelevanceassessments} genuinely supports their claim, particularly when the test collection is intended to serve as a benchmark for future research innovations.
Second, we submit a system deliberately crafted to exploit automatic evaluation metrics, demonstrating that it can achieve artificially inflated scores without truly improving retrieval quality.
Third, we simulate the consequences of circularity by analyzing Kendall’s tau correlations under the hypothetical scenario in which all systems adopt {\sc Umbrela} as a final-stage re-ranker, illustrating how reliance on LLM-based assessments can distort system rankings.
Theoretical challenges~---~including the inherent narcissism of LLMs, the risk of overfitting to LLM-based metrics, and the potential degradation of future LLM performance~---~that must be addressed before LLM-based relevance assessments can be considered a viable replacement for human judgments.

\end{abstract}

\maketitle

\section{Introduction}

In early 2023, \citet{ictir23} applied
GPT-3.5 (\texttt{text-davinci-003}) to fully reassess runs submitted to the
TREC 2021 Deep Learning track.
They reported Kendall's $\tau = 0.86$ for human vs. LLM-based assessment
on NDCG@10.
A natural conclusion might be that LLMs could now replace humans for
routine relevance assessment.
Instead, \citet{ictir23} issue a warning.
While recognizing the potential of LLMs to improve ranking and acknowledging their value as part of the relevance assessment process,
they argue strongly against abandoning human assessment.
They raise concerns about the potential for unknown biases that LLM-based assessments might introduce.
They highlight the issue of circularity, where LLMs evaluate the outputs of other LLMs.
Most importantly, their primary concern is that ``LLMs are not people.'' 
Since information retrieval systems are designed to serve human needs, their evaluation must ultimately reflect human judgment and preferences.

Recently, \citet{upadhyay2024largescalestudyrelevanceassessments} analyze the retrieval task results from the TREC 2024 RAG track. This retrieval task (or ``R task'') mirrors a traditional TREC ad hoc retrieval task. Participating systems were tasked with executing 301 queries over the MS MARCO Segment V2.1 collection, producing a ranked set of 100 passages for each query (a ``run''). They compare four procedures for assessing runs:
1) fully automatic assessments using the {\sc Umbrela} LLM-based relevance assessment
tool~\cite{upadhyay2024umbrelaumbrelaopensourcereproduction};
2) fully manual assessments using established TREC evaluation protocols for human judgments;
3) a hybrid method where {\sc Umbrela} filtered the set  of passages to be judged; and
4) a hybrid method where humans refined {\sc Umbrela}’s assessments.
In this paper, we focus on the first two procedures:
fully automatic and fully manual assessments. Based on their analysis,
\citet{upadhyay2024largescalestudyrelevanceassessments} conclude:
\begin{quote}
\textit{
Our results suggest that \textbf{automatically generated {\sc Umbrela} judgments
can replace fully manual judgments} to accurately capture run-level
effectiveness. Surprisingly, we find that LLM assistance does not
appear to increase correlation with fully manual assessments, suggesting that \textbf{costs associated with human-in-the-loop processes
do not bring obvious tangible benefits}....
Our work validates the use of LLMs in academic TREC-style
evaluations and provides the foundation for future studies.
}    
\end{quote}
\textbf{We disagree.} Not only does their reported correlation fail to provide stronger evidence than that of \citet{ictir23}, but additional evidence from the track directly contradicts their conclusion.
This evidence includes runs submitted by \anoni{one of the authors' research teams (WaterlooClarke)}{team WaterlooClarke}, which were explicitly designed to subvert LLM-based relevance judgments by employing LLM-generated judgments as a final-stage ranker.

\citet{ictir23} already demonstrated a strong empirical correlation between manual judgments and LLM judgments, both in terms of inter-annotator agreement and leaderboard correlation, and many other work also observed this empirical correlation.  However, after a detailed consideration of competing views, \citet{ictir23} concluded that there are too many theoretical concerns before human judgments can be replaced. These concerns, which remain critical to the discussion, \textbf{have neither been addressed nor refuted} in the work of \citet{upadhyay2024largescalestudyrelevanceassessments,upadhyay2024umbrelaumbrelaopensourcereproduction}. 

Conceptually, there is no fundamental difference between an LLM-based relevance assessment and an LLM-based re-ranking method. Both predict an affinity score for a passage to be relevant for a given query.
In contrast, human relevance judgments are privileged precisely because they \textsl{are} created by humans, and only humans can provide a gold standard for the evaluation of usefulness.
While employing LLMs to train and implement rankers can lead to substantial performance gains, these improvements risk being illusory if they fail to reflect human judgments.
The observation that LLM-based relevance judgments closely mimic the outcomes of human relevance judgments suggests that these LLM assessments may themselves represent a strong ranking method, rather than a valid evaluation metric.

We acknowledge the value of work by \citet{upadhyay2024umbrelaumbrelaopensourcereproduction}. In particular, their use of a new collection with fresh queries, which guarantees that LLMs were not trained on this collection. Moreover, their work confirms that the correlation seen by \citet{ictir23} is not merely an artifact of training on the test collection.
However, the warnings issued by \citet{ictir23} remain both valid and increasingly urgent, especially given the growing prevalence of LLM-based relevance assessments in information retrieval tasks.

\begin{figure}
\centering
\begin{overpic}[width=0.8\columnwidth]{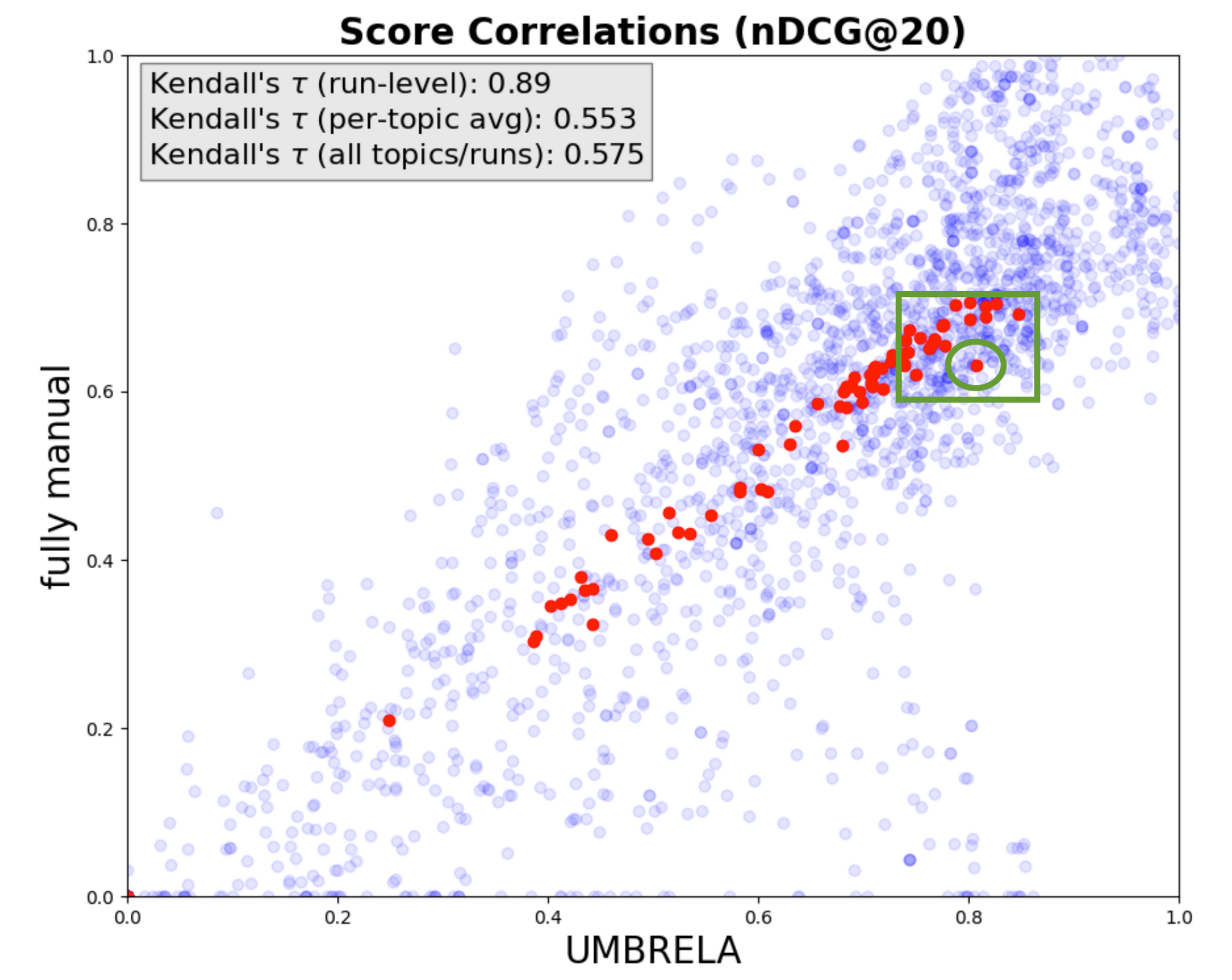} 
  \put(85,30){\includegraphics[width=0.2\columnwidth]{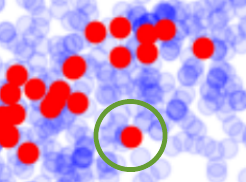}}
\end{overpic}
\caption{
Scatter plot extracted from Figure~3 of \citet{upadhyay2024largescalestudyrelevanceassessments} comparing manual assessment and automatic assessment.
Each red dot represents the average performance of a run over all queries.
The blue dots plot all run-query combination.
The inset provides a closer view of the points in the green rectangle.
}
\label{fig:avm}
\end{figure}

\section{Reproduction of Umbrela Results}
Kendall's $\tau$ is a suitable metric for showing overall
rank correlations on large-scale experiments.  \citet{upadhyay2024largescalestudyrelevanceassessments}
report a high overall Kendall's $\tau = 0.89$ between manual relevance assessments and automatic  {\sc Umbrela} assessments.
However, we note that some submitted systems perform substantially worse than others, making them easy to distinguish by an evaluation.  The presence of such under-performing systems can artificially inflate Kendall’s $\tau$ scores.

\begin{figure}
    \centering
    \includegraphics[width=1\linewidth]{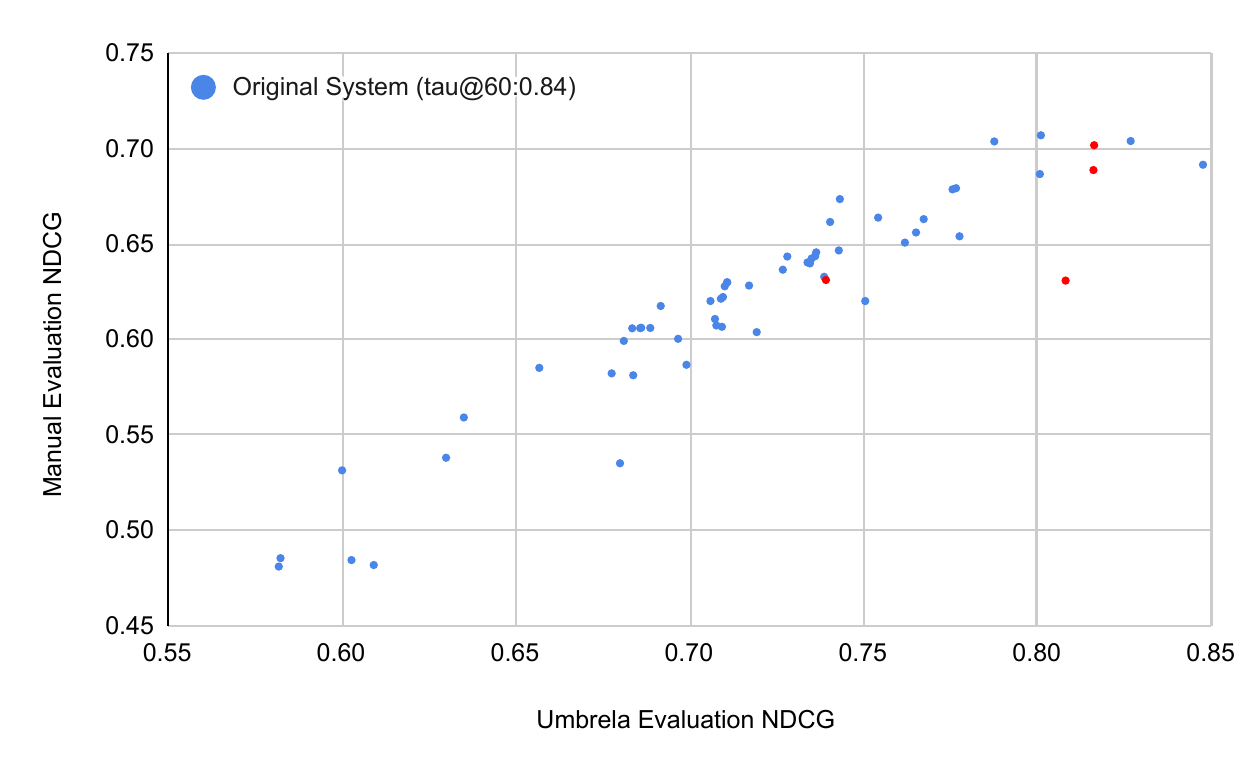}
    \caption{Reproduction of \citet{upadhyay2024largescalestudyrelevanceassessments}: On top 60 original TREC RAG 24 systems and data, the {\sc Umbrela} LLM evaluator correlates highly with manual assessors. Only few submitted retrieval systems included approaches from LLM evaluators.  Each
    system represents one dot. Red dots mark systems provided by team WaterlooClarke, which are known to contain LLM evaluations for re-ranking.}
    \label{fig:rag24-evaluator}
\end{figure}

To investigate this effect, we extend the original analysis by excluding the bottom-ranked 15 systems (out of 75) and recomputing Kendall’s $\tau$ over the remaining top 60 systems. As shown in Figure~\ref{fig:rag24-evaluator}, this yields a slightly lower but still relatively high Kendall’s $\tau$ of 0.84. This corresponds to approximately 8\% of system swaps, where the two evaluators disagree on which system performs better. Overall, these findings broadly confirm the results reported by \citet{upadhyay2024largescalestudyrelevanceassessments}.
Nevertheless, we observe some notable outliers. For example, one system achieves a high LLM-based evaluation score (0.81) but a substantially lower manual evaluation score (0.63), as shown in Figure~\ref{fig:rag24-evaluator}. We analyze such discrepancies further in Sections~\ref{sec:empirical-demonstration-of-the-risk} and~\ref{sec:llm-based-relevance-assessment-for-re-ranking}.

\section{Differences among Top-Performing Systems}
Demonstrating methodological advancements often involves re\-using test collections with the goal of surpassing state-of-the-art systems. Identifying meaningful differences among top-performing runs is therefore critical for measuring significant progress.
In contrast to the overall Kendall’s $\tau$ of 0.89 between manual and automatic assessments, the correlation weakens substantially among the highest-scoring systems.
While Kendall’s $\tau$ among the top 60 systems\footnote{Throughout the paper, when we refer to ``top-performing systems'', we use the manual relevance judgments as a basis for this assessment.} remains relatively high at 0.85 (corresponding to 8\% system swaps), it drops to 0.51 (24\% swaps) among the top 20 systems, and to 0.56 (21\% swaps) among the top 15.

Thus, when using an LLM-based evaluator to demonstrate improvements over state-of-the-art systems, precise agreement with manual judgments is essential at the top of the leaderboard.
A Kendall’s $\tau@15$ of 0.56 suggests that automatic {\sc Umbrela} assessments fail to demonstrate strong alignment with manual judgments at the top of the leaderboard. This misalignment undermines the reliability of automatic evaluations for tracking progress at the frontier of retrieval effectiveness.

Closer consideration of results from \citet{upadhyay2024largescalestudyrelevanceassessments} further highlights specific discrepancies in top-performing system rankings. 
Figure~\ref{fig:avm} reproduces a scatter plot extracted from Figure~3
of that paper, showing the performance of submitted runs (red dots)
under manual assessment (y-axis) vs.\ automatic assessment (x-axis).\footnote{
At the time of writing,
\citet{upadhyay2024largescalestudyrelevanceassessments}
do not provide a full public data release. They have generously provided limited access to their data for the purpose of confirming factual statements in this paper.
} 
Focusing on the top-performing systems (inset of Figure\ref{fig:avm}), discrepancies become evident:
for instance, the system ranked highest under automatic evaluation would only place fifth under manual evaluation. Conversely, the top system under manual evaluation ranks sixth under automatic evaluation. Particularly interesting is the case of the run circled in green. While it ranks fifth under automatic evaluation, it drops to 28th under manual evaluation.

These inconsistencies underscore a fundamental limitation of LLM-based assessments, such as those used by {\sc Umbrela}, in reliably identifying the best-performing systems. As a result, caution is warranted when using these methods for evaluating and validating progress in retrieval tasks.
While some differences could be the attributed to statistical
noise~---~only 27~queries were manually judged, and no error bars are provided~---~the available evidence remains insufficient to justify the replacement of human judgments.
This concern is particularly important when a collection is intended for re-use: claims of a novel system’s superiority over existing methods must be supported by improvements that are both statistically meaningful and aligned with manual assessments. At present, such confidence is lacking.

\begin{figure}
    \centering
    \includegraphics[width=1\linewidth]{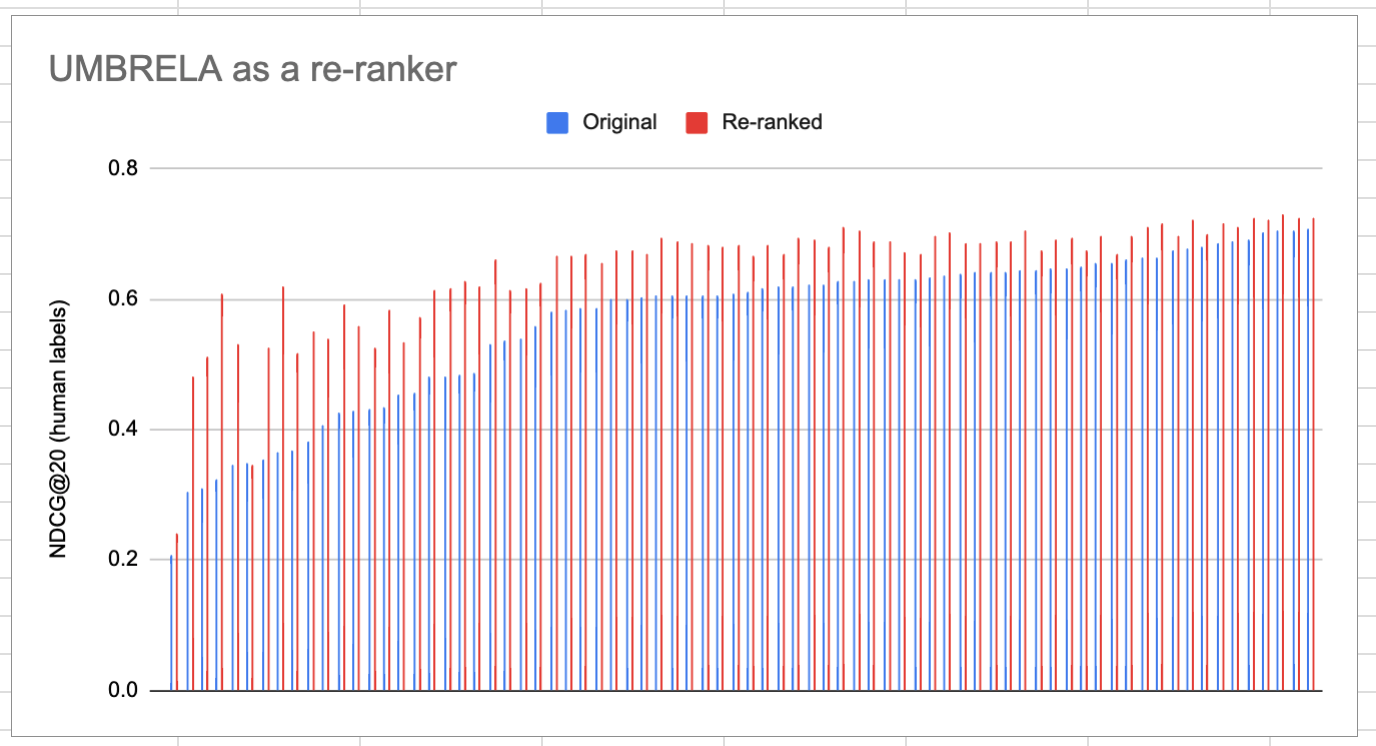}
    \caption{Reranking with an LLM evaluator ({\sc Umbrela}) improves performance under human relevance labels. This plot compares the original and reranked versions of all TREC RAG 24 systems based on manual assessment.}

    \label{fig:rag24-llmranker}
\end{figure}

\section{Subverting Automatic Evaluation}

When task relevance labels are generated entirely through a publicly known automatic process, such as {\sc Umbrela}, the evaluation metric becomes vulnerable to manipulation.
For instance, a participant could aggregate the outputs of many rankers, apply the {\sc Umbrela} system to this pooled set, and submit the resulting relevance labels as a new system for evaluation.
Such a strategy could, in principle, achieve perfect scores across all metrics. Even if the specific LLM-based relevance assessment process includes undisclosed elements, such as the exact prompt or LLM used, participants could approximate the process enough to subvert the automatic evaluations.

\begin{figure}
    \centering
    \includegraphics[width=1\linewidth]{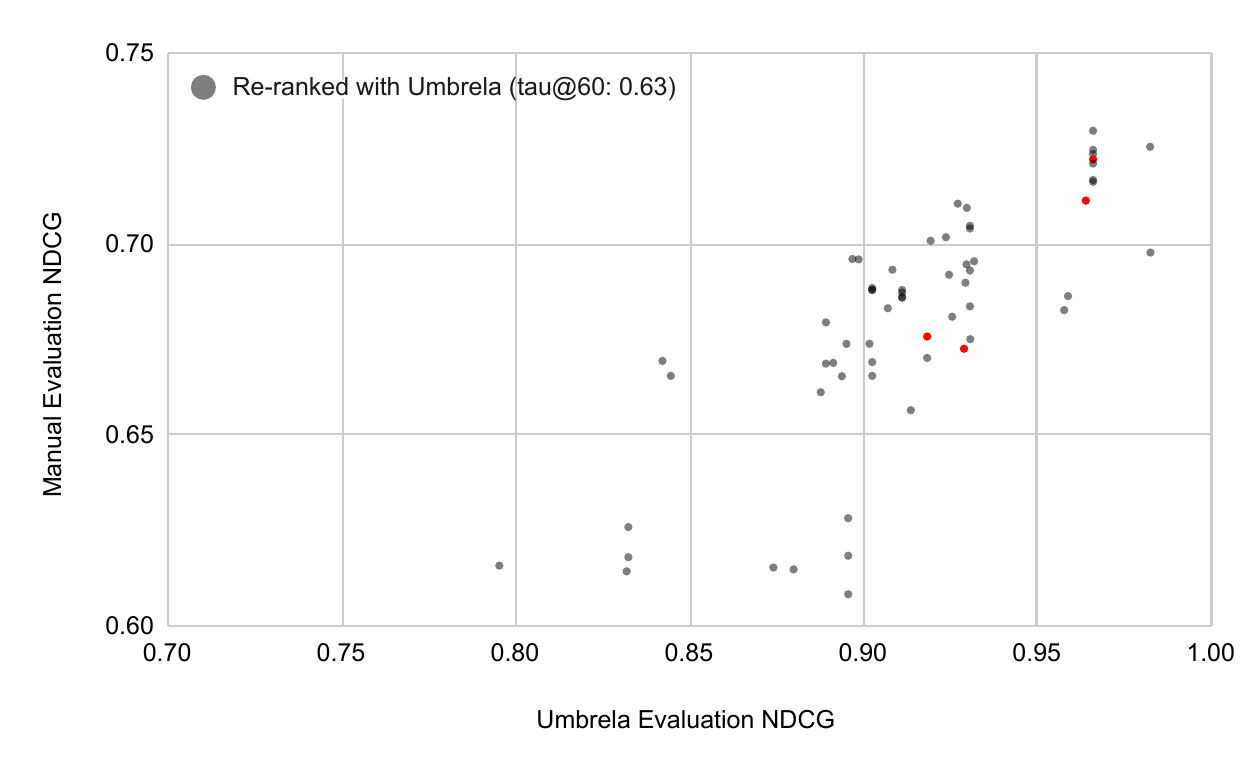}
    \caption{Demonstration of the effects of circularity when using {\sc Umbrela} as both evluator and ranker using TREC RAG 24 data. Each submitted retrieval
    system is first re-ranked with {\sc Umbrela}, then evaluated under NDCG with
    relevance labels from human judges and the {\sc Umbrela} evaluator. We see that
    especially among top ranked systems, the evaluation strategy no longer agrees
    with human judges on which system is better. Axis ranges are adjusted to display the same top 60 (of 75) systems as in Figure \ref{fig:rag24-evaluator}.}
    \label{fig:rag24-llmranker-evaluator}
\end{figure}

\subsection{Empirical Demonstration of the Risk}
\label{sec:empirical-demonstration-of-the-risk}

The run circled in green in Figure~\ref{fig:avm} exemplifies this vulnerability.
Submitted by \anoni{Clarke’s research group (\texttt{WaterlooClarke})}{team \texttt{WaterlooClarke}} as run \texttt{uwc1}, this submission was deliberately designed to subvert the automatic evaluation process.
Specifically, the team pooled the top 20~documents from 15 preliminary runs, spanning neural and traditional rankers, with and without query expansion.
This pool was then judged by \texttt{GPT-4o} with the prompt described by \citet{ictir23}.
Top-graded passages were subsequently judged pairwise following the procedure of \citet{pref} but substituting LLM-based assessments for crowdsourced human judgments.

The final ranking for \texttt{uwc1} was determined using LLM-based preference judgments as the primary key, LLM-based relevance assessments as the secondary key, and the reciprocal rank fusion~\cite{rrf} of preliminary runs as a tertiary tie-breaker. As intended, this deliberate attempt to manipulate the evaluation process led \texttt{uwc1} to rank significantly higher under automatic assessment (5th) than under manual assessment (28th).

\subsection{LLM-based Relevance Assessment for Re-Ranking}
\label{sec:llm-based-relevance-assessment-for-re-ranking}

As \citet{Soboroff_2025} recognizes: ``Retrieval and evaluation are the same problem. Asking a computer to decide if a document is relevant is no different than using a computer to  retrieve  documents  and  rank  them  in  order  of  predicted  degree  of  relevance.''
In this sense, LLM-based relevance assessment can be viewed as a specific form of LLM-based re-ranking.
To employ an LLM-based relevance assessment tool as a LLM-based re-ranker,
one starts with an initial ranking and prompts the LLM to assign a
score~---~expressed as a relevance grade~---~to each of the top-$k$ passages.
These passages are then re-ranked according to their assigned relevance grades, preserving their original order among passages with equal grades. 

This re-ranking interpretation is further illustrated by another run submitted by \anoni{Clarke’s research group (\texttt{WaterlooClarke})}{team \texttt{WaterlooClarke}}, \texttt{uwc2}. For \texttt{uwc2}, the team re-ranked the track's baseline run using the prompt described by \citet{genir}.
This re-ranking improves the baseline's performance from 9th to 4th place under manual assessments, and from 7th to 3rd place under automatic assessments.

\subsection{Simulation of Circularity}
\label{sec:circularity}

To explore the potential effects of widespread adoption, we simulate what would have happened if TREC RAG 2024 systems had incorporated {\sc Umbrela} as part of their pipeline.
Following the methodology described in Section~\ref{sec:llm-based-relevance-assessment-for-re-ranking}, we apply {\sc Umbrela} as a final-stage re-ranker to the outputs of all submitted retrieval systems.
For this experiment, we re-rank using the {\sc Umbrela} relevance labels from the track itself.
While the track organizers employed {\sc Umbrela} for evaluation, the labels it produced could instead have been employed as a final-stage ranker.
While we use the labels from the track organizers, any track participant could just as easily have employed {\sc Umbrela} as a final-stage re-ranker themselves.

Comparing original and re-ranked systems on manual (i.e., human) judgments in Figure~\ref{fig:rag24-llmranker}, we see that re-ranking with the {\sc Umbrela}
judgments consistently improves system performance.
TREC and similar evaluation experiments generally allow participants to use any automatic re-ranking process for their submissions. 
Since {\sc Umbrela} re-ranking is an entirely automatic process, it immediately loses its value as a measurement tool for those ranking experiments.
If any automatic re-ranking process can be used, using the measurement tool itself provides an optimal re-ranking~\cite{Soboroff_2025}.
More generally, if system performance is measured solely by LLM-based tools, system developers are strongly incentivized to incorporate the same tools into their systems.


Once we adopt {\sc Umbrela} as both a system component and an evaluation metric, it leads to an invalid \textbf{circular evaluation}.
This effect is demonstrated in Figure~\ref{fig:rag24-llmranker-evaluator}, where the {\sc Umbrela}-re-ranked runs from Figure~\ref{fig:rag24-llmranker} are evaluated with the same {\sc Umbrela} relevance labels used for re-ranking and compared against evaluations based on human judgments.
We observe a substantial increase in disagreement between the two evaluation methods, with discordant system pairs rising to 18\% within the top 60 systems.
As a result, Kendall’s $\tau$ drops sharply to 0.63.
This degradation becomes even more pronounced at the top of the leaderboard:
Kendall’s $\tau$ further decreases to 0.44 among the top 20 systems, 0.49 among the top 15, 0.38 among the top 10, and even turns negative ($\tau = -0.40$) among the top 5 systems.
Under {\sc Umbrela}-based evaluation, twelve systems now obtain NDCG scores exceeding 0.95, implying near-perfect ranking performance.
Yet the same systems achieve manual NDCG scores only between 0.68 and 0.72, illustrating substantial score inflation due to circularity.

For publication in peer-reviewed information retrieval research venues it is often necessary to demonstrate that a proposed system significantly outperforms all strong baselines.
Under a circular evaluation, however, such findings would no longer be credible.
These results highlight the risks of using LLM-based evaluation pipelines without safeguards against feedback loops, especially when test collections are intended for reuse.
Taking manually created relevance labels as the gold standard, we conclude that evaluating systems using the {\sc Umbrela} LLM assessor~---~when those systems internally apply {\sc Umbrela}-based re-ranking~---~results in an invalid circular experimental evaluation.

\section{Automatic Judgments are not Gold Standards}

The prompts used to elicit relevance grades from LLM-based assessment tools resemble instructions typically given to human assessors.  However, this resemblance is superficial and we should not be fooled by it. Such prompts merely represent one of many possible ways an LLM-based re-ranking method might assign scores, akin to an a LLM-based point-wise ranker. Despite being commonly referred to as relevance assessments, these scores are not equivalent to the judgments produced by humans.

LLM-based relevance assessments cannot serve as a gold standard because they lack the grounding of a human carrying out an information task necessary to evaluate the usefulness of retrieval systems. A true gold standard must originate from human assessments, as only humans can determine the relevance of information in a way that reflects real-world utility.

\citet{ictir23} raised concerns about the potential unknown biases inherent in LLM-based assessments. However, one clear and concerning bias is that LLM-based relevance assessments tend to favor LLM-based ranking systems.
Recently, \citet{balog2025rankersjudgesassistantsunderstanding} report a detailed evaluation of how LLM-based rankers can influence LLM-based judges, providing the first  empirical evidence that LLM judges exhibit ``a clear and substantial bias in favor of LLM-based rankers.''
This bias has been observed in other contexts as well \citep{liu2023narcissistic,wang2023unfair,panickssery2024own}, where LLMs demonstrate a form of ``narcissism,'' disproportionately favoring outputs generated by similar models.
Furthermore, \citet{alaofi2024llms} show that LLMs can be deceived through well-crafted prompt attacks embedded in content, leading them to incorrectly judge irrelevant text as relevant. These vulnerabilities highlight not only the susceptibility of LLM-based assessments to manipulation but also their inability to objectively evaluate diverse ranking approaches. Such biases and flaws further undermine the reliability of LLM-based assessments as a substitute for human judgments in critical tasks.

\section{When Automatic Judgments become Useless}
\label{sec:useless}

While the \texttt{uwc1} run demonstrates how a bad actor can strategically subvert an evaluation experiment, it is reasonable to assume that most participants are well-intentioned. 
These participants are not merely competing to win but are contributing to the creation of reusable test collections that support the development of innovative systems. However, even without malicious intent, the next generation of information systems will likely incorporate the latest advancements in LLMs, including prompting LLMs for relevance. As a result, some ranking methods will inherently embed elements that mirror LLM-based relevance judgments.
In a future evaluation experiment, it is plausible that even a well-intentioned participant could inadvertently undermine the evaluation process.

Looking ahead, we anticipate that future retrieval systems will increasingly rely on automatically generated training data to optimize machine learning components. Here, Goodhart's law serves as a cautionary principle \cite{goodhart1975problems}: ``When a measure becomes a target, it ceases to be a good measure.'' While the current observed correlation between manual and automatic assessment methods is strong, we predict that this correlation will degrade as developers incorporate LLM-based evaluation components into their systems in more refined ways than our simulation in Section \ref{sec:circularity}. Over time, these systems risk becoming disconnected from the human judgments they are intended to serve.
If the entire end-to-end experimental pipeline~---~from query formulation to relevance labeling~---~is fully automated, the evaluation process devolves into an LLM assessing its own assessments. The circularity feared by \citet{ictir23} is no longer a hypothetical concern; it has already begun to manifest in practice.

\section{Conclusion}

This paper raises serious concerns about the claims made by \citet{upadhyay2024largescalestudyrelevanceassessments}, which presents a preliminary analysis of data from the retrieval (``R'') task of the TREC 2024 RAG Track.
The author list of \citet{upadhyay2024largescalestudyrelevanceassessments} includes some of the most prominent experts in the area of information retrieval evaluation.
Despite being preliminary, their conclusions strongly imply that LLM-based relevance assessment can replace human relevance assessment~---~a claim that does \textbf{not} withstand scrutiny.
Given the authority of the authors and the strength of their implied conclusions, there is a risk that these findings may gain widespread acceptance within the research community without sufficient critical consideration. 
Nearly two years ago, \citet{ictir23} reached the opposite conclusion based on similar evidence.
Their concerns have still not been addressed.

\section*{Acknowledgments}

We have discussed our concerns with some of the authors of \citet{upadhyay2024largescalestudyrelevanceassessments}, and we appreciate their attention and feedback.
They also generously provided us with early access to their data to confirm factual statements in this paper.

This material is based in part upon work supported by the National Science Foundation under Grant No. 1846017. Any opinions, findings, and conclusions or recommendations expressed in this material
are those of the author(s) and do not necessarily reflect the views of the National Science Foundation.

\balance
\bibliographystyle{ACM-Reference-Format}
\bibliography{paper}

\end{document}